# Reliability and Validity of the Polar V800 Sports Watch for Estimating Vertical Jump Height

**Manuel V. Garnacho-Castaño** [1]✉, **Marcos Faundez-Zanuy** [2], **Noemí Serra-Payá** [1], **José L. Maté-Muñoz** [3], **Josep López-Xarbau** [2] and **Moisés Vila-Blanch** [4]

[1] GRI-AFIRS, School of Health Sciences, Pompeu Fabra University, Barcelona, Spain; [2] Polytechnic School Pompeu Fabra University, Barcelona, Spain; [3] Department of Physical Activity and Sports Science, Alfonso X El Sabio University, Madrid, Spain; [4] Department of Physical Activity and Sports Science, Higher Education Centre Palma de Mallorca, Spain

**Abstract**

This study aimed to assess the reliability and validity of the Polar V800 to measure vertical jump height. Twenty-two physically active healthy men (age: 22.89 ± 4.23 years; body mass: 70.74 ± 8.04 kg; height: 1.74 ± 0.76 m) were recruited for the study. The reliability was evaluated by comparing measurements acquired by the Polar V800 in two identical testing sessions one week apart. Validity was assessed by comparing measurements simultaneously obtained using a force platform (gold standard), high-speed camera and the Polar V800 during squat jump (SJ) and countermovement jump (CMJ) tests. In the test-retest reliability, high intraclass correlation coefficients (ICCs) were observed (mean: 0.90, SJ and CMJ) in the Polar V800. There was no significant systematic bias ± random errors ($p > 0.05$) between test-retest. Low coefficients of variation (<5%) were detected in both jumps in the Polar V800. In the validity assessment, similar jump height was detected among devices ($p > 0.05$). There was almost perfect agreement between the Polar V800 compared to a force platform for the SJ and CMJ tests (Mean ICCs = 0.95; no systematic bias ± random errors in SJ mean: -0.38 ± 2.10 cm, $p > 0.05$). Mean ICC between the Polar V800 versus high-speed camera was 0.91 for the SJ and CMJ tests, however, a significant systematic bias ± random error (0.97 ± 2.60 cm; $p = 0.01$) was detected in CMJ test. The Polar V800 offers valid, compared to force platform, and reliable information about vertical jump height performance in physically active healthy young men.

## Introduction

Vertical jumping is one of the most used assessments for measuring and monitoring explosive strength in several sports (Castagna and Castellini, 2013; García-Pinillos et al., 2014; Ferioli et al., 2018; Ulbricht et al., 2016). From a metabolic and biomechanical perspective, jump height is usually used for estimating the capacity and power of anaerobic metabolism (Bosco et al., 1983; Dal Pupo et al., 2014), as well as to assess the mechanical and neuromuscular fatigue induced by different types of exercise in the lower-body (Garnacho-Castaño et al., 2015; 2019; Gathercole et al., 2015). Furthermore, vertical jump testing has been considered a useful method to evaluate physical fitness in various populations such as children (Acero et al., 2011), highly-trained athletes (Carroll et al., 2019), healthy young adults (Garnacho-Castaño et al., 2018a; Maté-Muñoz et al., 2017) or elderly people (Pereira et al., 2012). It has also been used to evaluate the success of treatment methods or post-operative protocols in the clinical setting (Petschnig et al., 1998).

Various methods have been described in the scientific literature to measure vertical jump ability (Balsalobre-Fernández et al., 2014; Bosco et al., 1983; Glatthorn et al. 2011; García-López et al., 2005). Maximum jump height is estimated by applying the appropriate equations for vertical velocity at take-off and impulse (García-López et al., 2005). Currently, due to the plethora of new simpler and cheaper devices, flight time has been considered the most common way to estimate vertical jump height and has been shown to be valid and reliable (Balsalobre-Fernández et al., 2014; Bosco et al., 1983, Dias et al., 2011; Glatthorn et al. 2011). Flight time is determined as the period between take-off and contact after flight.

Diverse apparatuses and protocols have been proposed to assess vertical jump height. A force platform is considered as the "gold standard" for measuring jump ability (Glatthorn et al. 2011) and the vertical velocity of the center of mass at take-off (TOV) can be calculated by integrating the vertical force trace to measure vertical jump height. Video analysis systems, which measure displacement of the center of gravity of the body from the standing position to the highest vertical height, have been proposed as criterion instrument in several studies (Leard et al., 2007; Requena et al. 2012). This system has demonstrated excellent accuracy (García-López et al., 2005; Leard et al., 2007) for measuring flight time. Both devices are relatively expensive and require qualified personnel for the use of the material and specific software. In most cases, these evaluations are not specific to the sports environment and are generally restricted to laboratory-based settings (Walsh et al., 2006).

Portable and cost-effective devices have been designed to assess the vertical jump ability in the same conditions as in sports training occur. In this regard, photoelectric cells (Glatthorn et al., 2011), contact platforms (Loturco et al., 2017), accelerometers (Casartelli et al., 2010), high-speed camera systems (Balsalobre-Fernández et al., 2014) and several apps (Balsalobre-Fernández et al., 2015; Bogataj et al., 2020) analyzing the flight time have been used for assessing jump height with validity and reliability. Despite the important technological advance, the functionality and portability of these apparatuses could be improved in other sports environments. These devices are not normally used in fitness centers or in amateur sports for assessing physical fitness performance and require prior knowledge for software management and data interpretation.

Heart rate monitors and sport watches are considered one of the most used tools in sports training and in the field of physical fitness for training and assessment purposes (Borresen and Lambert, 2008; Garnacho-Castaño et al., 2018b). This is because a sports watch is easy-to-wear, inexpensive and unobtrusive tool (Conraads et al., 2012). The Polar V800 accelerometer has been validated to measure "1 hour sedentary bouts" and "lifestyle time" in young adults in free-living (Hernández-Vicente et al., 2016). Another function of this device is that the Polar V800/Stride Sensor system (Polar V800) can measure the vertical jump ability using a sensor that is placed in the sport shoes. The Polar stride sensor system is a tri-axial acceleration sensor for measuring flight time. This sensor telemetrically sends the time-of-flight records to the watch on the wrist, which estimates the height of the vertical jump. However, the reliability and validity of this device to estimate vertical jump height has not been explored.

This study aimed to assess the reliability and validity of the Polar V800 to measure vertical jump height. We hypothesised that the Polar V800 sports watch is a reliable and valid tool to measure jumping ability. This instrument would enable the measurement of jump height in addition to its other, more traditional, functions.

## Methods

### Participants

Twenty-two healthy young men were recruited for the study (age: 22.89 ± 4.23 years; body mass: 70.74 ± 8.04 kg; height: 1.74 ± 0.76 m; body mass index: 23.14 ± 2.01). All participants were students from the School of Health Sciences (Sport Sciences and Physiotherapy) that engaged in physical activity at least three times per week. All of the participants were considered healthy and injury free and were completely familiarized with the squat jump (SJ) and the countermovement jump (CMJ).

Before the tests commenced, participants were informed of the purpose of the study and the experimental procedures, and written consent was obtained from each subject. The study was approved by the Ethics Committee of the University according to the principles and policies of the Declaration of Helsinki.

### Experimental design

The reliability and validity of the Polar V800 was assessed by comparing jump height measurements simultaneously obtained using the force platform (gold standard) and the high-speed camera in two identical sessions (test-retest) one week apart. Both jump modalities were carried out according to the guidelines established in a previous study (Maté-Muñoz et al., 2014). Participants began with a general warm-up, which consisted of 5 min in cycle-ergometer followed by 5 min of dynamic stretches and joint movements of the arms, legs and trunk.

Next, participants completed a specific warm-up composed of three SJ and CMJ. The rest time between each jump was 30 seconds. After 3 min passive rest, participants started with the test protocol that involved 3 SJ and 3 CMJ (6 jumps per participant in each session) with a 1 min recovery period between each jump. Each jump was executed on a force platform while simultaneously recording the vertical jump height with a Polar V800 and a high-speed camera (Figure 1). Participants were asked to refrain from physical effort, smoking or the intake of caffeine, alcohol or nutritional supplementation 24 hours before each testing session. All tests were carried out at the same time of day (± 1 hour) and under similar environmental conditions (20°C-24°C and 60%-75% humidity).

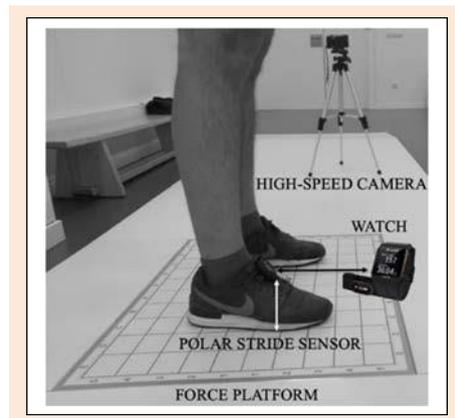

**Figure 1.** Distribution and location of the Polar stride sensor Bluetooth smart, high-speed camera and force platform during squat and countermovement jumps

### Jumping technique

The SJ test began from an initial position with hips and knees flexed (~90°) avoiding countermovement and maintaining this position for about 4 s to avoid the build-up of elastic energy stored during flexion to be used by leg extensor muscles. From the position of the hips and knees flexed ~90°, a knee and hip extension was performed as rapidly and explosively as possible.

The CMJ test started from a static standing position with hands on their hips. From this position, the subject underwent a rapid flexion-extension of the knees and hips. As in SJ, the knee joint was at an angle of around 90° of knee flexion. The depth of the SJ and CMJ was controlled in the high-speed camera by a researcher. The squat jump was checked using the high-speed camera and on the force platform (forces before take-off) to ensure that the participant did not perform a countermovement. Each jump was checked by two experienced researchers. The jumps that were not executed with an adequate technique were not considered for the statistical analysis.

For both jumps, the hands were placed on hips to avoid any help with the arms. During the flight phase, the knees and hips were completely extended. The landing was completed concurrently with both feet keeping the legs and hips extended until contact was made with the force platform. Participants were asked to take off and land at the same location to avoid lateral or horizontal displacement. After contact was made with the force platform, participants were allowed to control the landing by flexing their hips, knees and ankles. Before the start of each trial, participants were instructed to jump as high as possible.

**Equipment and data acquisition**
To check the validity of the Polar V800 (Polar Electro OY; Kempele, Finland), a force platform (Musclelab, Ergotest Technology AS, Langesund, Norway) was used as a "gold standard" device. In addition, a high-speed EX-FC100 camera (Casio Computer Co., Ltd Exilim, Tokyo, Japan) was used to determine possible differences with the Polar V800 sports watch.

The Polar V800 (37 mm x 56 mm x 13 mm; mass 79 gr) is operated by a 350 mAh Li-pol rechargeable battery. The Polar Stride Sensor Bluetooth Smart (Polar Electro OY; Kempele, Finlandia) is a tri-axial acceleration sensor responsible for measuring flight time at a sampling rate of 100 Hz. This sensor was placed on sports shoes as indicated in figure 1. Before the jump test began, the sensor was synchronized by Bluetooth with the watch situated on the wrist. Once the jump was executed, this sensor sent telemetrically the flight time records to the watch placed on the wrist, which estimated jump height by means of the following equation [5]: $h = g \cdot ft^2 / 8$; h = jump height in meters; g = acceleration due to gravity ($m \cdot s^{-2}$); ft = flight time in seconds. Finally, jump height was recorded on the watch in centimeters.

The force platform (60cm x 40cm x 7cm; mass 12.7 kg) recorded data at a sampling frequency of 1 kHz. The force sensors of this platform were constituted by 4 strain gauge 5kN, total max 20kN. The force platform was connected to a portable computer with the specific software (Ergotest Technology AS, Langesund, Norway).

The high-speed camera (99.8 mm x 58.5 mm x 22.6 mm; 145 gr) filmed each jump at a frequency of 240 frames per second with a shutter speed of 1/4000. The high-speed camera recordings were successively analyzed using video analysis Kinovea (software 0.8.7. for windows). The Kinovea software calculated flight time of each jump by identifying the take-off and the landing frames in the video analysis, and then transforming it into a jump height using the equation $h = g \cdot ft^2 / 8$ [5], as in the Polar V800.

During the recording of the jump, the camera was placed on a tripod perpendicular to the sagittal plane of the participants and at a distance of 3 meters and 0.90 meters high. Markers were placed on the greater trochanter, femoral epicondyle and lateral malleolus (Tsoukos et al., 2016) of each subject as a reference for further analysis using the Kinovea software. The force platform, the high-speed camera and the Polar V800 measured the flight times simultaneously during SJ and CMJ.

**Statistical analysis**
All statistical tests and graphics were performed using SPSS software version 25.0 for Mackintosh (SPSS, Chicago, IL). The Shapiro–Wilk test was used to check the normal distribution of data, provided as means and standard deviation (SD). To identify significant differences between test-retest in the Polar V800 (reliability) and between the Polar V800, force platform and high-speed camera/Kinovea system (HSC/KS) (validity), a general linear model with a two-way analysis of variance (ANOVA) for repeated measures was carried out. The two factors were exercise mode x time (3 devices x test-retest). A Bonferroni post hoc adjustment was used to test for differences among pairs of means (multiples comparisons). Partial eta-squared ($\eta_p^2$) was computed to determine the magnitude of the differences. The statistical power (SP) was also calculated.

The relative reliability (test-retest) was evaluated with ICC (2, 1) with a 95% CI. The absolute reliability was examined using Bland Altman systematic bias ± random error and the coefficient of variation (CV), expressed as a percentage of the mean results, and was calculated as the typical error of measurements (Bland and Altman, 1986). The concurrent validity was determined using the intra-class correlation coefficient (ICC) (2, 1) with a 95% confidence interval (CI) and Bland-Altman method systematic bias ± random error (Bland and Altman, 1986). The scale for classification of ICC was 0.90 = very good, and good = between 0.71 and 0.90 (Bartko, 1966). The range for classification of CV was 3.1%-8.6% (Hopkins et al., 2001). Proportional bias was assessed by linear regression between the averages and the differences in the results obtained in Bland-Altman plots. Significance was set at $p \leq 0.05$.

**Results**

No significant exercise mode x time interaction effects or time effect were detected in SJ and CMJ tests (p > 0.05). Only an exercise mode effect was observed in CMJ test (p < 0.001; $F_{(2, 94)} = 10.39$, $\eta_p^2 = 0.18$, SP = 0.99). After Bonferroni comparison, similar vertical jump height was found between the test-retest in the Polar V800 (reliability) (p > 0.05) in SJ and CMJ tests (Figure 2). Significant higher jump height values were observed in force platform than in HSC/KS (p = 0.001) in CMJ test (Figure 2). Similar jump height values were observed in Polar V800 with regards to force platform and HSC/KS (validity) in SJ and CMJ tests (p > 0.05).

**Test-retest reliability**
The range of ICCs was between 0.83 and 0.90 in SJ and CMJ (Table 1). The Bland-Altman plots indicated that there was no significant systematic bias ± random errors (mean: -0.04 ± 3.18 cm in SJ; -0.39 ± 3.51 cm in CMJ) (p > 0.05) between test-retest in all devices. There was no evidence of proportional bias between test-retest (Figure 3). Low CVs were detected in SJ (5.05%) and CMJ (5.27%) tests (Table 1).

**Concurrent validity**
An almost perfect agreement was found (ICCs mean: 0.90 in SJ and 0.95 in CMJ) (Table 2). No significant systematic bias ± random errors were observed among devices in SJ (mean: -0.13 ± 2.71 cm in SJ, p > 0.05). However, a significant systematic bias ± random errors was detected between HSC/KS versus Polar V800 (p = 0.01) and force platform (p < 0.001) in CMJ test (Table 2). No proportional bias was found among devices in SJ and CMJ tests (Figure 4).

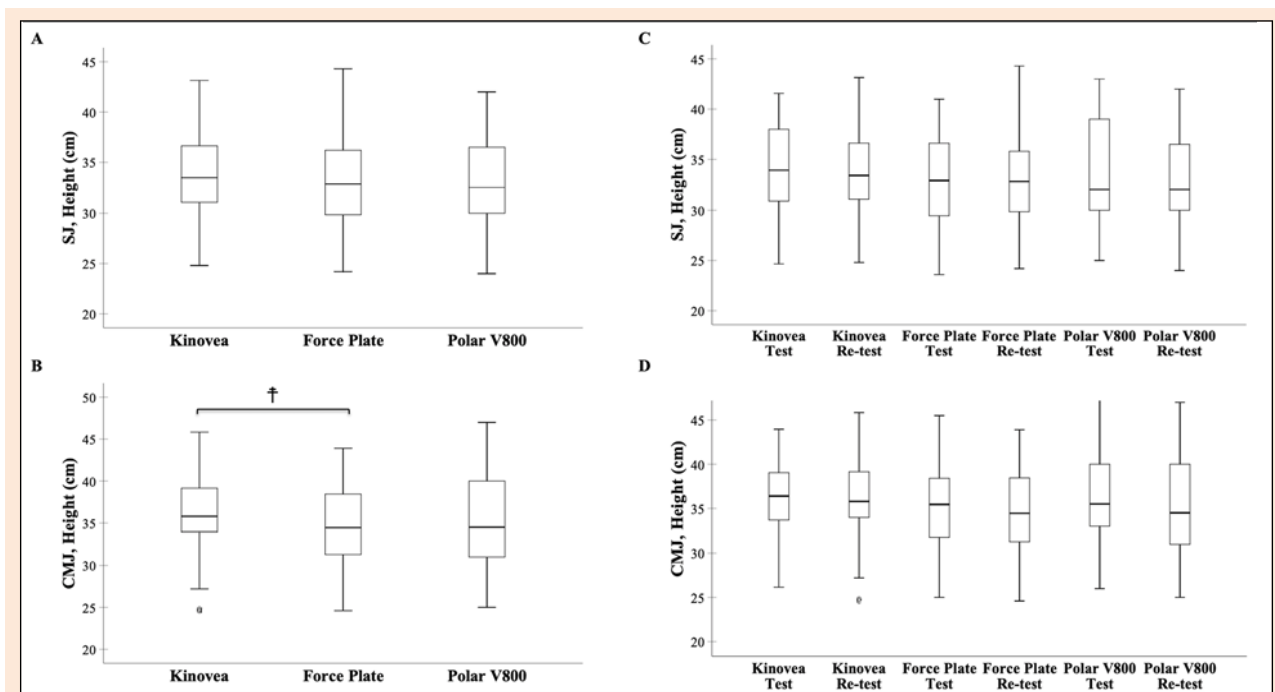

**Figure 2.** Vertical jump height values acquired: **(A)** Squat jump (SJ) in force platform, high speed camera/ Kinovea system (HSC/KS), and the Polar V800/Stride Sensor. **(B)** Countermovement jump (CMJ) in force platform, HSC/KS, and the Polar V800/Stride Sensor. **(C)** SJ in test-retest assessment in the three devices. **(D)** CMJ in test-retest assessment in the three devices.
\* significant differences, p = 0.001.

**Table 1.** Test-retest reliability of the Polar V800 sports watch, force platform and high-speed camera

|                  | Systematic Bias (cm) | Random Error (cm) | Proportional Bias | ICC (95% CI)     | CV (%) |
|------------------|----------------------|-------------------|-------------------|------------------|--------|
| Polar V800, SJ   | -0.37                | 3.14              | No                | 0.90 (0.80-0.93) | 4.83   |
| Force Plate, SJ  | 0.44                 | 3.59              | No                | 0.83 (0.71-0.90) | 6.20   |
| Kinovea, SJ      | -0.18                | 2.81              | No                | 0.89 (0.81-0.94) | 4.11   |
| Polar V800, CMJ  | -0.77                | 3.29              | No                | 0.90 (0.81-0.94) | 4.66   |
| Force Plate, CMJ | -0.09                | 3.82              | No                | 0.84 (0.73-0.91) | 5.90   |
| Kinovea, CMJ     | -0.31                | 3.44              | No                | 0.86 (0.77-0.92) | 5.24   |

CI: confidence interval; CMJ: countermovement jump; CV: coefficient of variation; ICC = intraclass correlation coefficient; SJ: squat jump. No significant test-retest differences in each device, p > 0.05.

**Table 2.** Concurrent validity of the Polar V800 sports watch, force platform and high-speed camera.

|                            | Systematic Bias (cm) | Random Error (cm) | Proportional Bias | ICC (95% CI)     |
|----------------------------|----------------------|-------------------|-------------------|------------------|
| FP vs. Polar V800, SJ      | -0.30                | 2.36              | No                | 0.93 (0.88-0.96) |
| HSC/KS vs. Polar V800, SJ  | 0.69                 | 2.70              | No                | 0.90 (0.83-0.94) |
| FP vs. HSC/KS, SJ          | -0.79                | 3.07              | No                | 0.88 (0.79-0.93) |
| FP vs. Polar V800, CMJ     | -0.45                | 1.85              | No                | 0.97 (0.94-0.98) |
| HSC/KS vs. Polar V800, CMJ | 0.97*                | 2.60              | No                | 0.93 (0.88-0.96) |
| FP vs. HSC/KS, CMJ         | -1.30*               | 2.19              | No                | 0.96 (0.92-0.98) |

CI: confidence interval; CMJ: countermovement jump; FP = force platform; HSC/KS = high-speed camera/Kinovea system; ICC = intraclass correlation coefficient; SJ: squat jump. *Significant differences among devices, p ≤ 0.01.

## Discussion

The purpose of this study was to evaluate the reliability and concurrent validity of the Polar V800/Stride sensor system for measuring vertical jump height performance. A very good test-retest reliability was observed for the estimation of jump height in the Polar V800. In addition, the Polar V800 demonstrated very good concurrent validity compared with force platform (gold standard) and HSC/KS in SJ and CMJ tests.

In the test-retest assessment, although a very good agreement (relative reliability) was found in SJ and CMJ test (ICCs = 0.90) in the Polar V800, we recorded slightly lower ICCs to those reported in previous studies that assessed vertical jump height (Balsalobre-Fernández et al., 2015; Glatthorn et al., 2011). Also, slightly lower ICCs were recorded in force platform and HSC/KS (Mean ICCs ~0.86). ICCs above 0.90 are considered as very good, and values between 0.71 and 0.90 as good (Bartko, 1966).

The CVs detected in all devices (range: 4.11% to 6.20%) suggest adequate absolute reliability in both jump test modalities according to the range (3.1%-8.6%) reported by Hopkins et al. (2001). The range of the CVs stated by Hopkins et al. (2001) was obtained in measurements on force platforms, contact mats and yardsticks.

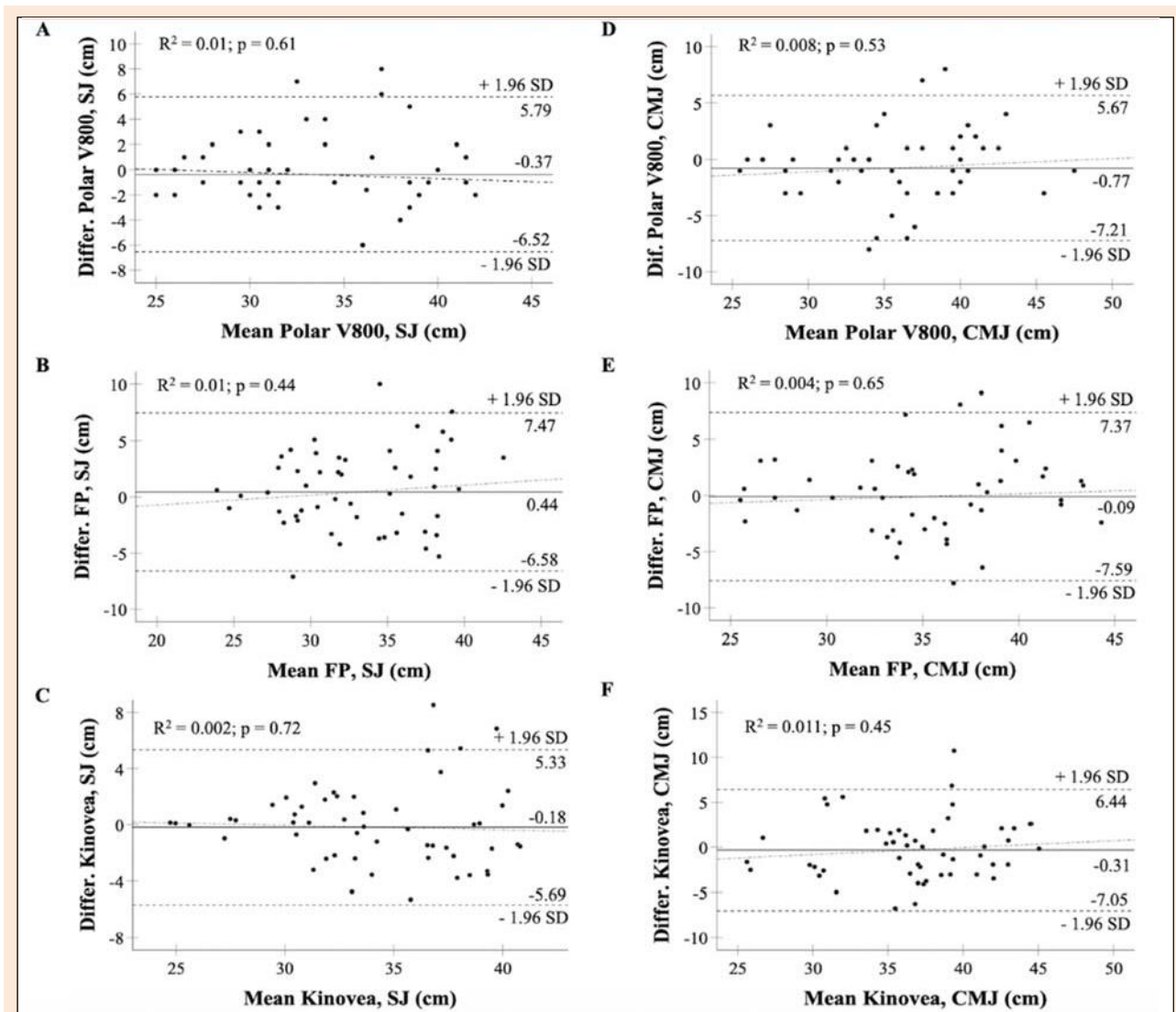

**Figure 3. The Bland Altman plots in the reliability (test-retest) assessment: (A) the Polar V800/Stride Sensor in squat jump (SJ). (B) Force platform in SJ. (C) High-speed camera/ Kinovea system (HSC/KS) in SJ. (D) The Polar V800/Stride Sensor in countermovement jump (CMJ). (E) Force platform in CMJ. (F) HSC/KS in CMJ.** $R^2$ refers to proportional bias. Dif. or differ. = differences.

On the whole, negligible systematic differences ± random errors were detected between session 1 and 2 (test-retest) with values of 1.32% for the Polar V800, 0.39% for force platform and 0.71% for HSC/KS (n > 100 jumps in each device). In test-retest reliability assessment, the consistency or stability of measurements observed in this study is a key factor to guarantee that detected variances between testing sessions in vertical jump height are not produced by a systematic bias, such as fatigue induced or learning influence, or random error due to mechanical or biological variations (Atkinson and Nevill, 1998). It is likely that the accuracy of the reliability measurements was due to the fact that all participants were highly familiarized with the SJ and CMJ technique.

The strong concurrent validity shown by the Polar V800 was based on the comparison in the vertical jump height with another criterion instrument that takes into consideration the same variables to be tested. Despite differences among devices in sampling frequency, similar jump height values were observed between the Polar V800, the force platform (gold standard or criterion instrument) and the HSC/KS (2nd reference system) in SJ and CMJ tests. In addition, the Bland Altman plots revealed insignificant systematic bias ($p > 0.05$) between the Polar V800 and force platform. The average difference in vertical jump height among both devices was 0.88% in SJ and 1.32% in CMJ (< 0.5 cm, fig. 4A y 4D). Also, no systematic bias ($p > 0.05$) was observed between the Polar V800 and the HSC/KS. The average difference in jump height among both apparatuses was 1.77% in SJ (0.69 cm, fig. 4B).

For strengthening this validity assessment, a linear regression between the averages and the differences in Bland-Altman plots showed that no proportional biases were found between the Polar V800 and both reference systems after having performed more than 100 jumps (included SJ and CMJ). Significant systematic biases (~1 cm) were found in similar studies that compared vertical jump height (SJ and CMJ) between various apparatuses with a force platform (gold standard) (Balsalobre-Fernández et al., 2015; Glatthorn et al., 2011). Glatthorn et al. (2011) confirmed that the difference in vertical jump height between an infrared platform and a force platform was approximately 2.5% or 1.06 cm. Analogous conclusions were drawn (1.1 ± 0.5 cm) when *My jump* application (high speed camera at 120 Hz) was compared with a force platform (1 kHz) (Balsalobre-Fernández et al., 2015). Accelerometer technology has shown a higher mean difference of 3.6 cm (CMJ) and 5.6 cm (SJ) in comparison with a force platform (1 kHz) (Choukou et al., 2014).

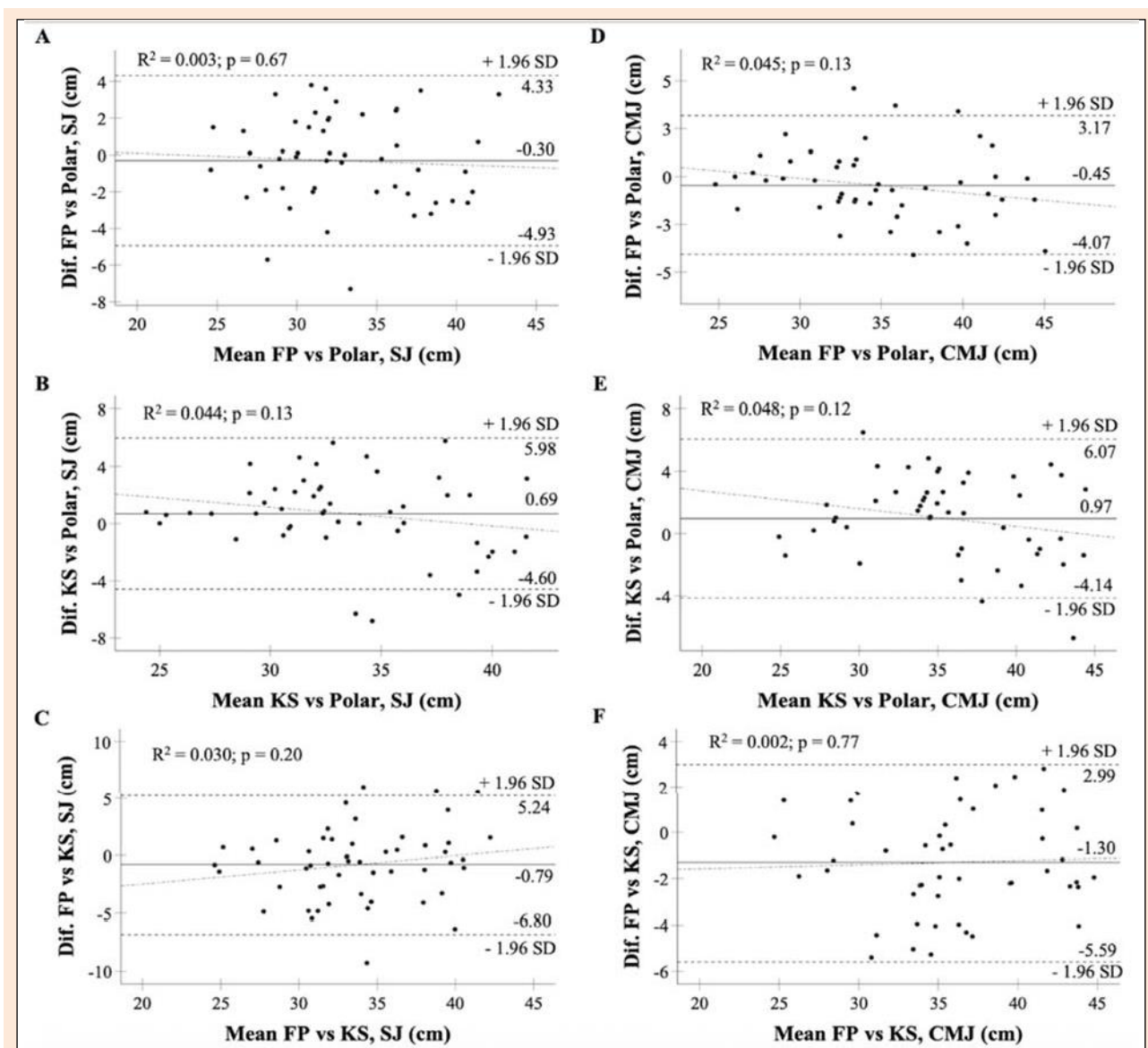

**Figure 4.** The Bland Altman plots in the validity assessment: **(A)** Force platform (FP) vs. the Polar V800/Stride Sensor in squat jump (SJ). **(B)** High-speed camera/ Kinovea system (HSC/KS) vs. the Polar V800/Stride Sensor in SJ. **(C)** FP vs. HSC/KS in SJ. **(D)** Force platform (FP) vs. the Polar V800/Stride Sensor in countermovement jump (CMJ). **(E)** HSC/KS vs. the Polar V800/Stride Sensor in CMJ. **(F)** FP vs. HSC/KS in CMJ. $R^2$ refers to proportional bias. Dif. = differences.

Previous studies have shown contradictory findings in relation to the sampling rate. Some authors determined that the differences observed in jump height were caused by the different sampling frequency observed among devices (Balsalobre-Fernández et al., 2015), however, other authors established that the differences in jump height were not due to the sampling rate (Glatthorn et al., 2011). The similar results obtained in this study by the three devices suggests that vertical jump height was not affected by the sampling rate.

In contrast, a slightly and significant systematic bias ± random error was detected in the Polar V800 and in the force platform regarding the HSC/KS in CMJ test. Requena et al. (2012) demonstrated that a high-speed camera recording data at 1 kHz produced a similar flight time (difference of 1.3 ms) to that of a force platform at the same sampling frequency. A high-speed camera at 120 Hz recorded data with a difference about 8.9 ms or 1.2 cm in comparison with a force platform (1 kHz) (Balsalobre-Fernández et al., 2015). In our study, the high-speed camera (240 Hz) showed a significant difference (p<0.05) in vertical jump height to those of a Polar V800 (2.27% or 0.97 cm) and a force platform (3.52% or 1.30 cm) at different sampling frequencies (100 Hz and 1 kHz, respectively) only in the CMJ test. When the technical execution of the jumps was analyzed in the Kinovea system, we suspected that CMJ could produce a greater variability than SJ during landing and air-time phases. It has been widely shown that a greater jump height is reached in CMJ than in SJ (Bobbert et al., 1996). The eccentric phase leads to a greater force at the start of the concentric phase. The increase in the force applied during the concentric phase could condition the take-off, the flight and the landing phases. Therefore, systems that measures flight time to estimate jump height might produce errors because the lift off and the landing positions are different (García-López, 2000; Kibele, 1998). The time elapsed between takeoff and landing during video analysis (for example, Kinovea) (e.g., Kinovea) could be different at the time elapsed when the force platform begins and finishes recording forces on platform surface (García-López et al., 2005).

Jump height values obtained with the Polar V800 should be based on criteria of consistency or agreement with one or more measuring devices. In this case, a variance analysis providing ICCs is recommended, which analyze inter-subject and inter-observer variability and the residual error (Bartko, 1966). High ICCs were detected between the Polar V800 and both the force platform (mean ICCs = 0.95) and HSC/KS (mean ICCs = 0.92) in SJ and in CMJ tests.

The findings from this study suggest that the Polar V800 is a useful tool to measure vertical jump height in physically active young people who practice physical activity and recreational sport. The reliability and validity of this device is not necessarily extrapolated to the jumps of athletes with high leg power and other types of jumping (for example, Abalakov). In theory, the jump height calculated from the flight time would increase in jump modalities such as Abalakov. Probably, a wide range of participants (high performance athletes, healthy young men, adults, women, etc.) and mainly if the arms were included, would result in a much larger range of jump height values. Probably, these alterations in the range of the flight height values would affect the systematic and proportional bias. Studies evaluating the reliability and validity of the Polar V800 on vertical jump abilities in high performance athletes, other populations or other types of jumps are needed.

A key factor to consider in this study is that the number of participants may influence the understanding of the Bland Altman plots and the interpretation of statistical power of the data. Twenty-two subjects participated in this study; however, we suggest that the large number of jumps performed (> 100 per type of jump) are sufficient data to consider the Polar V800 sports watch a reliable device to measure the uniformity or absence of systematic and random errors.

Finally, sports watches are one of the most used tools in sports and fitness. The main findings suggest that the Polar V800 is a simple, multifunctional, inexpensive and practical device that offers adequate information about the vertical jump height performance in healthy young men. The Polar V800 appears to be a versatile and reliable tool for quantifying jump performance.

## Conclusion

The Polar V800 sports watch was shown to be a reliable and valid tool for measuring jumping ability. The low biases and random errors observed determine that the Polar V800 appears to be a useful device for assessment of vertical jump height in physically active young men.

### Acknowledgements
The authors have no conflict of interest to declare. The experiments comply with the current laws of the country in which they were performed. The datasets generated during and/or analyzed during the current study are not publicly available, but are available from the corresponding author who was an organizer of the study.

---

### Key points

- The Polar V800 sports watch was shown to be a reliable and valid tool for measuring jumping ability.
- The Polar V800 is a simple, multifunctional, inexpensive and practical device that offers adequate information about the vertical jump height performance in physically active healthy young men.
- It would also be interesting to demonstrate the reliability and validity of the Polar V800 sports watch to assess vertical jump height in highly trained athletes or other populations.

**AUTHOR BIOGRAPHY**

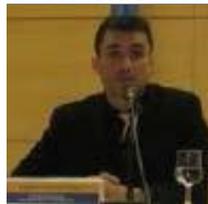

**Manuel Vicente GARNACHO-CASTAÑO**
**Employment**
Director of School of Health Sciences, TecnoCampus-Pompeu Fabra University, GRI-AFIRS, Barcelona, Spain.
**Degree**
PhD
**Research interests**
Exercise Physiology, Sports Performance.
**E-mail:**
mgarnacho@escs.tecnocampus.cat

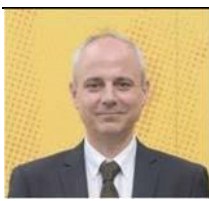
**Marcos FAUNDEZ-ZANUY**
**Employment**
Polytechnic School, TecnoCampus-Pompeu Fabra University, Barcelona, Spain.
**Degree**
PhD, Professor
**Research interests**
Engineering, new technologies applied to sport.
**E-mail:** faundez@tecnocampus.cat

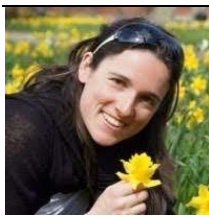
**Noemí SERRA-PAYÁ**
**Employment**
Deputy director of School of Health Sciences, TecnoCampus-Pompeu Fabra University, GRI-AFIRS, Barcelona, Spain.
**Degree**
PhD
**Research interests**
Exercise Physiology, Exercise prescription.
**E-mail:** nserra@tecnocampus.cat

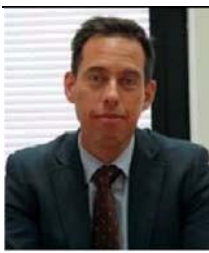
**José Luis MATÉ-MUÑOZ**
**Employment**
Head of Department of Physical Activity and Sport Sciences. Alfonso X el Sabio University, Madrid, Spain.
**Degree**
PhD
**Research interests**
Exercise Physiology, Resistance Training, Endurance Training.
**E-mail:** jmatmuo@uax.es

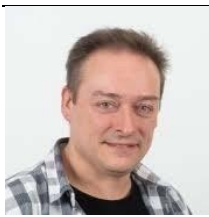
**Josep LÓPEZ-XARBAU**
**Employment**
Polytechnic School, TecnoCampus-Pompeu Fabra University, Barcelona, Spain.
**Degree**
Master's degree
**Research interests**
Engineering, new technologies applied to sport.
**E-mail:** jlopez@tecnocampus.cat

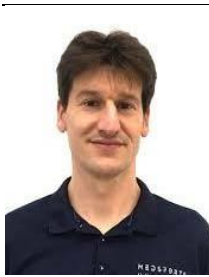
**Moisés VILA-BLANCH**
**Employment**
Associate professor, Department of Physical Activity and Sports Science, Higher Education Centre Alberta Gimenez, Universidad Pontificia Comillas, Palma de Mallorca. Spain
**Degree**
PhD
**Research interests**
Sports Performance
**E-mail:** moivilaperformance@gmail.com

✉ **Manuel V. Garnacho-Castaño**
TecnoCampus-Pompeu Fabra University Ernest Lluch, 32 (Porta Laietana), Mataró-Barcelona, 08302, Spain.